\def\kms{\relax \ifmmode {\,\rm km\,s}^{-1}\else \,km\,s$^{-1}$\fi}   
\def\nii{[N~{\sc ii}]}      
\def\oiii{[O~{\sc iii}]}      
\def\ha{H$\alpha$}    
\def\hb{H$\beta$}      
\def\stry{{Str\"omgren {\it y}}}    
\def\rr{{$r^\prime$}}     
\def\sexa{\object{Sextans~A}}    
\def\leoa{\object{Leo~A}}    
\def\ic10{\object{IC~10}}    
\def\sexb{\object{Sextans~B}}    
\def\hii{H~{\sc ii}}

\documentclass[onecolumn]{aa}    
\usepackage{graphicx}    
\begin{document}    
   \title{The Local Group Census:\\    
    planetary nebulae in \ic10, \leoa\ and \sexa\thanks    
{Based on observations obtained at the 2.5m~INT telescope      
operated on the island of La Palma by the Isaac Newton Group in the Spanish      
Observatorio del Roque de Los Muchachos of the Instituto de Astrofisica de      
Canarias.}}     
    
\author{     
L.     Magrini   \inst{1},       
R.L.M. Corradi   \inst{2},      
R.     Greimel   \inst{2},   
P.     Leisy     \inst{2,3}    
D.J.   Lennon    \inst{2},      
A.     Mampaso   \inst{3},    
M.     Perinotto \inst{1},      
D.L.   Pollacco  \inst{4},    
J.R.   Walsh     \inst{5}   
N.A.   Walton    \inst{6}, and      
A.A.   Zijlstra  \inst{7}  
}      
\offprints{R. Corradi\\       
e-mail: rcorradi@ing.iac.es}      
\institute{     
Dipartimento di Astronomia e Scienza dello Spazio, Universit\'a di     
Firenze, L.go E. Fermi 2, 50125 Firenze, Italy     
\and       
Isaac Newton Group of Telescopes, Apartado de Correos 321, 38700 Santa       
Cruz de La Palma, Canarias, Spain    
\and     
Instituto de Astrof\'{\i}sica de Canarias, c. V\'{\i}a L\'actea s/n,      
38200, La Laguna, Tenerife, Canarias, Spain     
\and     
School of Pure and Applied Physics, Queen's University Belfast, Belfast BT7     
9NN,    
Northern Ireland, UK    
\and       
ST-ECF,     
E.S.O., Karl-Schwarzschild-Strasse 2, 85748 Garching bei M\"{u}nchen, Germany     
\and      
Institute of Astronomy, University of Cambridge, Madingley Road, Cambridge   
CB3 0HA, UK 
\and      
Physics Department, UMIST, P.O. Box 88, Manchester M60 1QD, UK     
}

\date{Received ; accepted }    
   
\abstract{In the framework of our narrow-band survey of the Local    
Group galaxies, we present the results of the search for planetary 
nebulae (PNe) in the dwarf irregular galaxies 
\ic10, \leoa\ and \sexa.    
Using the standard on-band/off-band technique, sixteen new candidate
PNe have been discovered in the closest starburst galaxy, \ic10.  The
optical size of this galaxy is estimated to be much larger than
previously thought, considering the location of the new PNe in an area
of 3.6~kpc$\times$2.7~kpc.  We also confirm the results of previous
studies for the other two dwarf irregular galaxies, with the detection
of one candidate PN in \leoa\ and another one in \sexa.  We review the
number of planetary nebulae discovered in the Local Group to date and
their behaviour with metallicity. We suggest a possible fall in the
observed number of PNe when [Fe/H]$<<$-1.0, which might indicate that
below this point the formation rate of PNe is much lower than for
stellar populations of near Solar abundances.  We also find
non-negligible metallicity effects on the \oiii\ luminosity of the
brightest PN of a galaxy.
 }    
\authorrunning{Magrini et al.}      
\titlerunning{Planetary nebulae in \ic10, \leoa\ and \sexa}       
   \maketitle   
%
%
    
\section{Introduction}    
    
Most of the galaxies in the Local Group (LG) are dwarf irregulars and 
spheroidals.  These morphological types also represent the most 
numerous objects in the nearby universe, but can be studied in great 
detail only at the close distance of the LG. We are performing a 
narrow- and broad-band filter survey, the Local Group Census, which is mainly 
aimed at studying all classes of emission-line populations in the 
LG. The first results were presented by Magrini et 
al. (\cite{magrini02}, hereafter M02), Magrini et al.  
(\cite{magrini02b}), Wright et al. (\cite{wright02}), while the status 
of the project is described at {\it 
http://www.ing.iac.es/$\sim$rcorradi/LGC}. M02 presented the detection 
of planetary nebulae (PNe) in Sextans~B. Our next targets were the 
dwarf irregular galaxies \ic10\ (morphological type IrIV according to 
Van den Bergh \cite{vdbergh}, hereafter vdB00), \leoa\ (IrV) and 
\sexa\ (IrV), that we discuss in the present work.   
    
\ic10\ is a highly obscured galaxy (E(B-V)$\simeq$0.85, Sakai et    
al. \cite{sakai99}) located at a low Galactic latitude   
$b$$=$$-3^{\circ}.3$. Its distance (660 kpc, Sakai et al. \cite{sakai99})   
and its position (only $\simeq$18$^{\circ}$ apart from M~31 on the   
sky) suggest a possible membership to the M~31 subgroup (vdB00).  It   
is the only starburst galaxy in the LG, and the presence of a large   
number of \hii\ regions (Hodge \& Lee \cite{hodgelee}) proves that it   
is undergoing massive star formation.  \ic10\ is a rather small galaxy   
with an effective radius $r_e=$0.5 kpc (de Vaucouleurs \& Ables   
\cite{devau}), only one half the effective radius of the Small   
Magellanic Cloud (SMC), whereas their luminosities are comparable.
Its oxygen abundance is higher than that of SMC (12 + log(O/H)=8.20
compared with 7.98 in SMC; Skillman et al. 
\cite{skillman}), showing a higher past rate of star formation.  This
galaxy is clearly resolved in stars on ground-based images, and a
large number of Wolf-Rayet stars is known (Massey et
al. \cite{massey}).  The presence of a large foreground extinction due
to its location in a direction close to the Galactic plane has
prevented so far deep studies of the stellar populations. A first
search for PNe was undertaken by Jacoby \& Lesser (\cite{jacoby81},
hereafter JL81), and resulted in the discovery of one PN.
    
\leoa\ is a small irregular galaxy ($M$$<$$9\times10^7 M_{\sun}$, Young    
\& Lo \cite{young}) at a distance of 800 kpc (Dolphin et    
al. \cite{dolphin}), now firmly considered as a member of the LG
(vdB00). It contains both old and young population components (Tolstoy
et al. \cite{tolstoy}), the older one amounting to approximately 10\%
of the stellar population located near to the centre of the galaxy.
Strobel et al. (\cite{stroebel}) have detected several \hii\ regions
excited by hot stars and an unresolved emission-line object, probably
a planetary nebula. Heavy element abundance of this object shows a
very low metallicity ($\sim$2.4\%  solar; Skillman et al.
\cite{skillman}).  In this galaxy, JL81 discovered two candidate PNe.
    
The membership of the LG of \sexa, located at the distance of 1.45 Mpc
(Sakai et al.  \cite{sakai96}), is instead doubtful. It
could form a possible group with \object{NGC~3109}, \object{Antlia}
and \sexb\ (vdB00). \sexa\ seems to contain a conspicuous
intermediate-age population, as suggested by its prominent red giant
branch (Dohm-Palmer et al. \cite{dohm}). Star formation, at present,
is concentrated in a \hii\ region complex observed by Hodge et al. 
(\cite{hodge}).  Skillman et al. (\cite{skillman}) found an oxygen abundance $\sim$3\% 
of Solar.  A PN was identified by JL81.
    
In this paper we present \oiii\ and \ha+\nii\ continuum-subtracted 
images of these three galaxies.  These lead  to the discovery of 16 new 
candidate PNe in \ic10. In \leoa, the two candidate PNe found by JL81 
were shown instead to be normal stars, while the unresolved emission-line 
object seen by Strobel et al. (\cite{stroebel}) is confirmed as a 
possible PN.  We also detect a candidate PN in \sexa, confirming the 
detection of JL81.  Observations are described in Sect.~2. Data 
reduction and analysis are presented in Sect. 3.  In Sect. 4, we 
discuss the results and in Sect. 5 we review the present knowledge 
about PNe in the LG, discussing their behaviour with metallicity. 
Summary and conclusions are given is Sect. 6. 
    
\section{Observations}    
    
\sexa\ (\object{DDO~75},  10h\,11m\,00.5s -04d\,41m\,29s, J2000.0),     
\leoa\ (\object{DDO~69},  09h\,59m\,26.4s +30d\,44m\,47s, J2000.0) and    
\ic10 (00h\,20m\,23.2s +59d\,17m\,30s, J2000.0) were observed     
using the prime focus wide field camera (WFC) of the 2.54~m Isaac   
Newton Telescope (La Palma, Spain), on February 2001, February 2002   
and June 2002.  The detector of the WFC is composed of four thinned   
EEV CCDs with 2048$\times$4096 pixels each, with a pixel scale of   
0.$\farcs$33.  The large size of the field of view of the camera,   
$34\arcmin \times34\arcmin$, allows to cover each galaxy in a single 
WFC pointing.  The filters used are: \oiii\ (500.8/10.0~nm), \ha + \nii\   
(656.8/9.5), \stry\ (550.5/24.0), \rr\ (Sloan r, 624.0/134.7).  The   
\stry\ and \rr\ filters were used as off-band images for     
continuum subtraction of \oiii\ and \ha+\nii\ images, respectively.    
    
Each exposure was split into three sub-exposures.  The total exposure    
times were 3600~sec for \oiii\ and \ha+\nii\ for all galaxies.  We    
also exposed for 3600~sec in \stry, except for \sexa\ for which we    
exposed for 1800~sec. The exposure times in the \rr\ filter were    
1200~sec for \leoa\ and \sexa, and 1800~sec for \ic10.  The nights of    
February 2002 were not photometric, thus short exposures of    
\ic10\ and of the Galactic PN IC~5117 (Wright et al.  in preparation) 
were taken on June 2002 to allow absolute flux calibration.     
The nights of February 2001 were however photometric and several    
observations of the spectrophotometric standard stars: BD+33~2642 and    
G191-B2B (Oke \cite{oke}) were made each night.    
      
\section{Data reduction and analysis}    
    
Data reduction was done using IRAF \footnote{ IRAF is distributed by
the National Optical Astronomy Observatories, which is operated by the
Association of Universities for Research in Astronomy, Inc.  (AURA)
under cooperative agreement with the National Science Foundation}. The
frames were de-biased, flat-fielded, and linearity-corrected using the
ING WFC data-reduction pipeline (Irwin \& Lewis
\cite{irwin}).  Then we corrected for geometrical distortions and    
aligned all frames to the \oiii\ one.  The sky background, measured in 
regions far from the galaxies, was subtracted from each frame.  During 
the observations of \leoa\ the seeing was 1\farcs3 through all 
filters.  For \sexa, it was 0\farcs9 in the \ha+\nii\ and \stry, 
and 1\farcs1 in the \oiii\ and \rr\ filters.  For \ic10, it was 
1\farcs1 in the \ha+\nii\ and \stry\ and 1\farcs3 in the \oiii\ and \rr\ 
filters. 
    
We used the standard on-band/off-band technique ({\sl cf.} Magrini et 
al. \cite{magrini00}) to identify emission-line objects. The \rr\ and 
\stry\ frames, properly scaled, were subtracted from the \ha+\nii\ and    
\oiii\ frames, respectively.  The flux of the emission-line objects    
were computed in the continuum-subtracted frames, using the IRAF task 
APPHOT. For the \oiii\ line, fluxes were transformed to equivalent 
V-mag magnitudes using $m_{\rm [O~III]}=-2.5\log F_{\rm [O~III]}-13.74$ 
(Jacoby \cite{jacoby89}).
 
The errors of the \oiii\ and \ha+\nii\ fluxes of the candidate PNe in
\sexa\ and \leoa, including photon statistics, background and flux 
calibration uncertainties, amount to a few percent.  The fluxes of 
PNe in \ic10\ have typical errors of $\sim$10\% for PNe with 
$m_{\rm [O~III]}<24.0$, 15-20\% for $24.0<m_{\rm [O~III]}<25.0$ 
and of 30\% or 
more for the faintest PNe.  These errors take into account photon 
statistics and the error on the zero point (2-3\%) for the flux 
calibration. 
    
The instrumental magnitudes were calibrated by convolving the spectrum   
of the spectrophotometric standard star (Oke \cite{oke}) or the PN   
IC5117 (Wright et al. in preparation) with the response curve of each   
filter.   
    
The astrometric solutions were computed using the IRAF tasks CCMAP and 
CCTRAN, using the USNO A2.0 catalogue (Monet et al. \cite{monet}) for 
\ic10\ and the APM POSS1 for \sexa\ and \leoa.  The accuracy of all    
solutions is $\sim$0\farcs3 {\it r.m.s.}    
    
\section{Candidate planetary nebulae}    
   
Candidate PNe were identified using the same criterion as in Magrini   
et al. (\cite{magrini00}), i.e. as {\it spatially unresolved}   
emission-line objects detected in the \oiii\ and \ha+\nii\   
continuum-subtracted frames.  The candidate PNe are listed in Table 1,   
with their positions, their \oiii\ and   
\ha+\nii\ fluxes and the equivalent V-mag m$_{\rm [O~III]}$.  
Fluxes in the \oiii\ line at 
$\lambda$=500.7~nm were corrected for the contribution of the 
companion oxygen line at $\lambda$=495.9~nm.  This contribution varies 
depending on the different heliocentric radial velocities of the 
galaxies (-348 km/s for \ic10, 20 km/s for \leoa, 324 km/s for \sexa\ 
from the Nasa Extragalactic Database) and amounts to 6\%, 11\% and 
14\% respectively. 
  
\begin{table*}       
\caption{PN candidates in \ic10, \leoa\ and \sexa.        
\oiii500.7 and \ha+\nii\  observed fluxes are given in units of    
10$^{-16}$~erg~cm$^{-2}$~s$^{-1}$. \oiii\ fluxes are corrected for the
contribution of the companion oxygen line at 495.9~nm (see text).
$R_{corr}$=\oiii/(\ha+\nii) is computed after correcting the fluxes for the
reddening with E(B-V)=0.85 for \ic10, 0.015 for \sexa, 0.02 for
\leoa (vdB00), following Mathis' prescription (\cite{mathis}). }
\begin{center}    
\begin{tabular}{l r r r r r r}       
\hline 
\hline       
Identification & \multicolumn{2}{c}{R.A. (2000.0) Dec.} & $F_{\rm [OIII]}$ &     
$F_{\rm H\alpha}$ & m$_{\rm [O~III]}$ & $R_{corr}$  \\     
\hline       
\object{IC10 PN1} 	&0:18:44.42  & 59:17:47.7 &7.17 & 6.09 & 24.12&2.8\\        
\object{IC10 PN2} 	&0:19:04.32  & 59:17:04.8 &7.95 & 3.68 & 24.01&5.2\\      
\object{IC10 PN3} 	&0:19:13.60  & 59:14:43.9 &11.8 & 8.46 & 23.58&3.5\\      
\object{IC10 PN4} 	&0:20:06.73  & 59:15:21.5 &3.95 & 3.69  &24.76&2.1 \\  
\object{IC10 PN5} 	&0:20:17.26  & 59:15:52.5 &5.12 & 2.63  &24.49&4.9 \\      
\object{IC10 PN6} 	&0:20:21.14  & 59:21:26.3 &1.15 & 0.85 & 26.10&3.2\\      
\object{IC10 PN7}    &0:20:22.22  & 59:20:01.6  & 66.1 & 150. &21.71&1.1\\    
\object{IC10 PN8} 	&0:20:28.82  & 59:07:21.4 &13.2 & 8.68 & 23.46&3.7\\      
\object{IC10 PN9} 	&0:20:32.08  & 59:16:01.6 &13.5 & 7.67 & 23.43&4.4\\      
\object{IC10 PN10} 	&0:20:40.98  & 59:18:49.4 &14.8 & 7.01  &23.33&5.2 \\     
\object{IC10 PN11} 	&0:20:42.89  & 59:16:30.6 &13.5 & 8.65  &23.43&3.9 \\     
\object{IC10 PN12} 	&0:20:46.28  & 59:21:04.5 &3.45 & $<$1  &24.66&- \\        
\object{IC10 PN13} 	&0:20:49.09  & 59:18:32.9 &2.12 & 2.53 & 25.44&2.0\\      
\object{IC10 PN14} 	&0:20:50.96  & 59:20:23.6 &6.07 & 2.17 & 24.30&7.0\\      
\object{IC10 PN15} 	&0:21:09.87  & 59:16:58.9 &17.1 & 8.96  &23.18&4.7 \\     
\object{IC10 PN16} 	&0:21:11.23  & 59:16:11.2 &4.96 & 11.9  &24.52&1.1 \\     
\hline
\object{LEOA PN1} 	&9:59:31.66 & 30:45:28.0 &65.0 & 65.0  &21.73& 1.1\\     
\object{SEXA PN1} 	&10:11:01.58  & -4:41:22.9  &33.1 & 63.1  &22.46&0.5 \\     
\hline    
\end{tabular}       
\end{center}    
\end{table*}       
   
\begin{figure*}   
\centering 
\includegraphics[width=17cm]{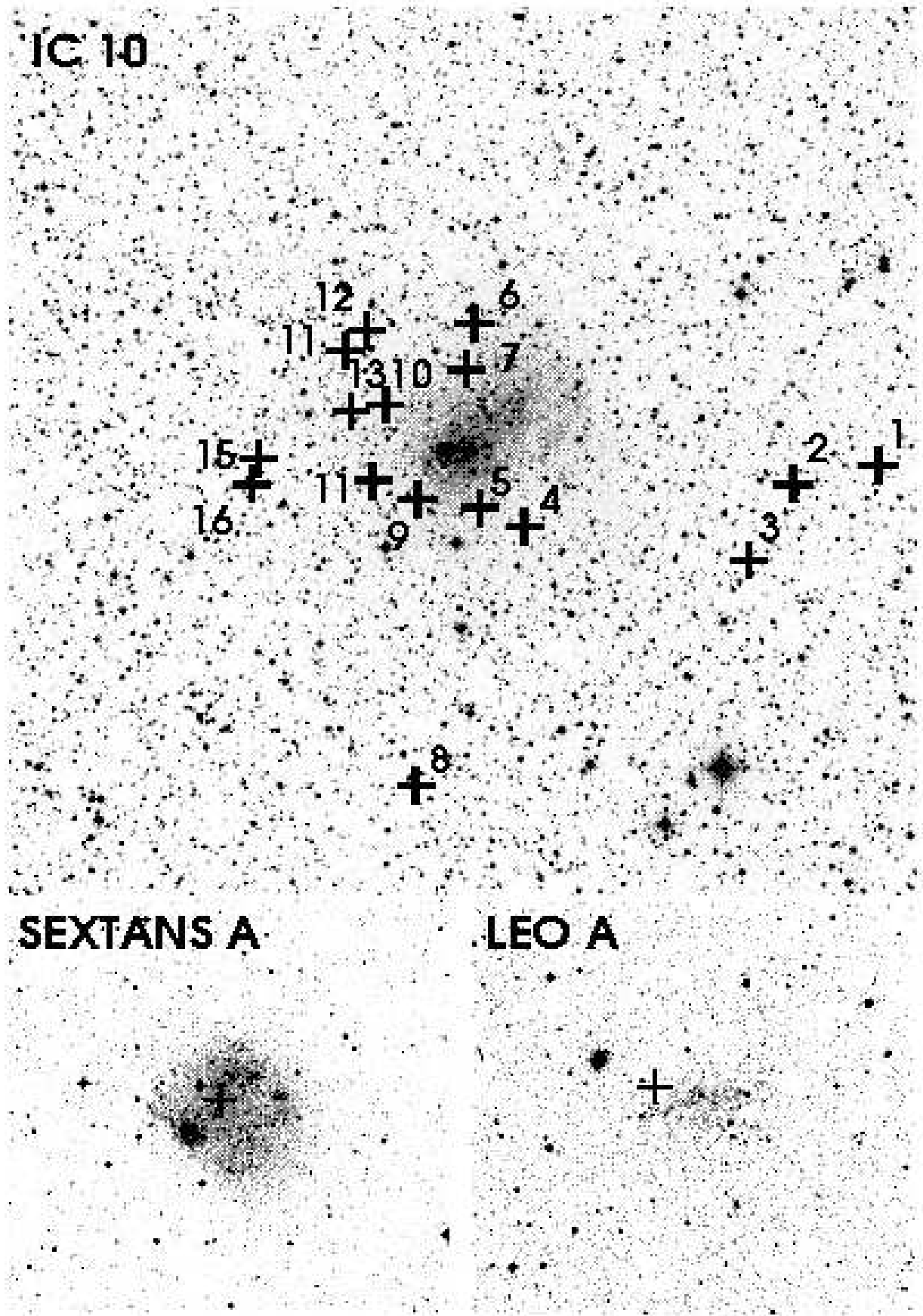}   
\caption{Digitized Sky Survey images of the galaxies \ic10,   
\leoa\ and \sexa. The size of the \ic10\ frame is   
30\arcmin$\times$30\arcmin, of \leoa\ and \sexa\ 
14\arcmin$\times$12\arcmin. North is at the top, East to the left. 
Candidate PNe are marked with a cross and their id. number as in 
Table~1.} 
\end{figure*}   
   
\subsection{Planetary nebulae in \ic10}    
    
Sixteen new candidate PNe were identified in the galaxy \ic10. This is
the first identification of PNe in \ic10, except for one PN discovered
by JL81, whose position was not reported.  Our candidate PNe are
detected both in the \oiii\ and \ha+\nii\ continuum-subtracted images,
so they cannot be confused with highly redshifted background
galaxies. For each PN, we compute the $R$=\oiii/(\ha+\nii) flux ratio
after correcting for a mean foreground extinction E(B-V)=0.85 (vdB00)
following Mathis (\cite{mathis}).  All candidate PNe have values of
the excitation index $R$ between 1.1 and 7.0, which is comparable with
values for Galactic PNe ({\sl cf.} Magrini et al. \cite{magrini00},
Fig. 3).  The PNe distribution looks spatially uniform, and no bias
against high/low $R$ or faint/bright PNe can be detected.
    
The PNe lie in a large area of 18\farcm7$\times$14\farcm1 (see
Figure~1), that corresponds to a linear size of 3.6~kpc$\times$2.7~kpc
for a distance of 660~kpc (Sakai et al. \cite{sakai99}).  No PNe are
found very close to the centre of this starburst galaxy, presumably
because of the presence of numerous extended \hii\ regions that cover a
large fraction of that area. Eight PNe are situated around the
starburst region, within 0.9~kpc to the centre, the closest ones (PN4
and PN8) being at $\sim$0.35~kpc.  The remaining PNe are located in
the outskirts of \ic10, outside the 25~mag~arcsec$^{-2}$ diameter
(5\farcm5$\times$7\farcm0, or 1.1$\times$1.3~kpc; Massey \&
Armandroff \cite{massey95}).  The farthest of our PNe (PN1) is at the
distance from the centre of 12\farcm7 (2.4~kpc).  On the other hand,
the large extent of the gaseous component of \ic10\ is well known.  
From 21-cm line observations, Huchtmeier (\cite{hu})
found the presence of an enormous neutral hydrogen envelope
(62\arcmin$\times$80\arcmin) surrounding the galaxy.  PNe have proven to
be excellent tracers of stellar populations in large volumes with a
relatively low density of stars, whose integrated stellar light is low
or even hardly detectable, like the intergalactic and intracluster
space and in the haloes of elliptical galaxies (Arnaboldi et
al. \cite{arna}).  Thus, if spectroscopic studies confirm the nature
as PNe of our candidates as well as their belonging to \ic10 (via
their radial velocities), they would reveal the presence of a
conspicuous stellar population at galactocentric distances much larger
than considered so far.  Moreover, Shostak \& Skillman
(\cite{shostak}) found that both the outer hydrogen envelope and the
core, which coincides with the small optical size of the galaxy, show
a rotation along the same axis, but with opposite sign.  It would be
extremely interesting to verify whether this kinematical dichotomy
also applies to PNe and to the stellar populations that they
represent.  

\begin{figure*}    
\resizebox{\hsize}{!}{\includegraphics{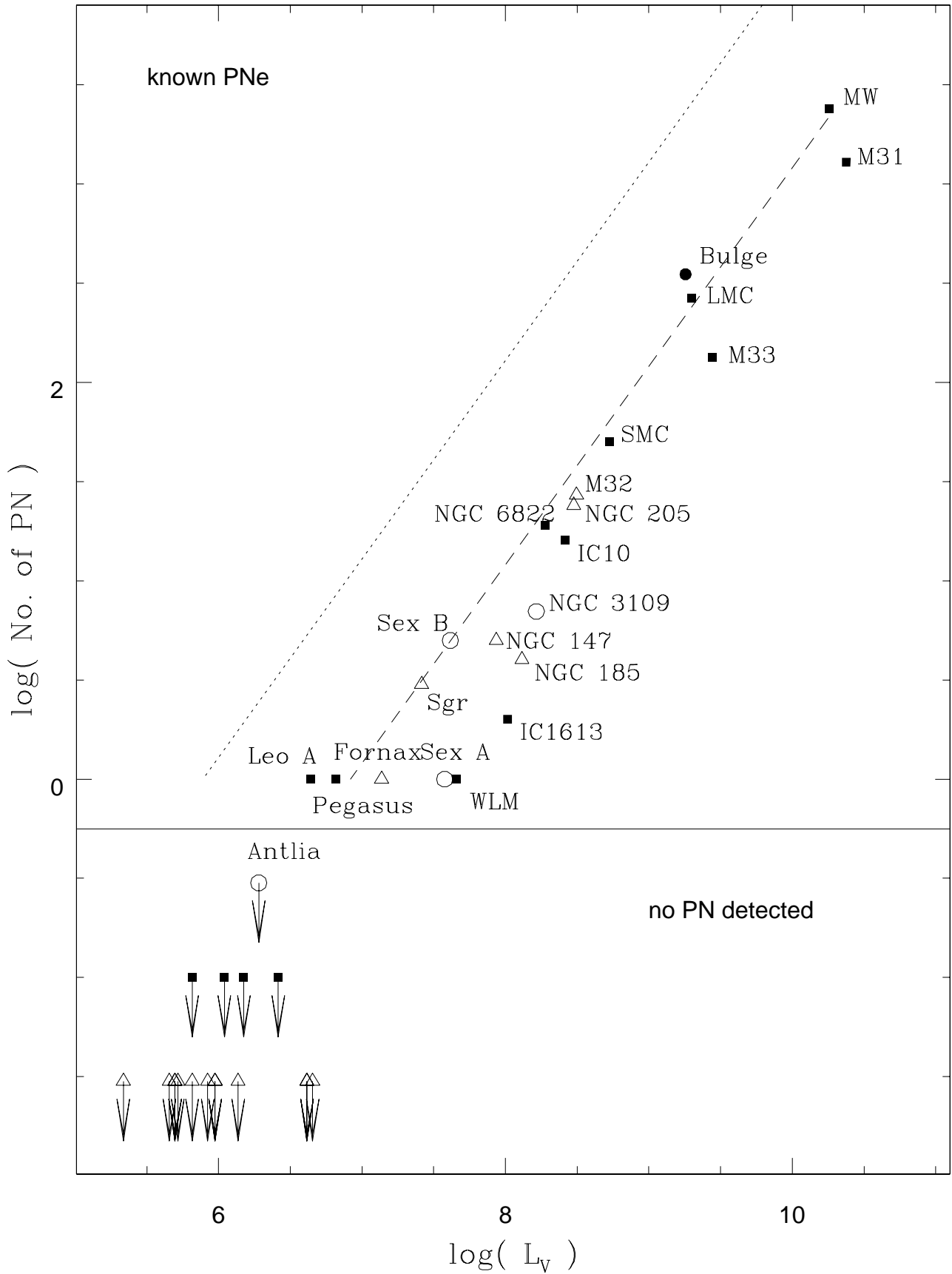}}  
      \caption{The number of PNe in galaxies in or near the Local 
      Group, versus the V-band luminosity in solar units.  The dotted line shows
the expected numbers based on a total number of PNe in the MW of
23000. The dashed line is fitted to the known population in the LMC.
Filled squares indicate LG gaseous galaxies, triangles indicate spheroidal
galaxies and open circles show the NGC 3109 group.  }  
\label{lgpn} 
\end{figure*} 
 
\subsubsection{Population size and completeness limit}    
    
The number of candidate PNe detected in \ic10\ is roughly consistent 
with the expected population size of this galaxy and its large 
foreground extinction (E(B-V)=0.85; vdB00).  The population size can 
be estimated using a model describing a simple (i.e coeval and 
chemically homogeneous) stellar population (Renzini \& Buzzoni 
\cite{renzini}).  As described by M02, the number of stars $n_j$ in   
any post-main-sequence phase $j$ is given by   
\begin{equation}      
n_{j}=\dot{\xi} L_T t_j, 
\end{equation}      
where $\dot{\xi}$ is the specific evolutionary flux (number of stars
per unit luminosity leaving the main sequence each year), $L_T$ the
total luminosity of the galaxy, and $t_j$ the duration of the
evolutionary phase $j$ ($\le$20~000 yrs for the PN phase).  Using the
values of the parameters as in M02, the corresponding population size
of \ic10\ is between 50 and 200 objects considering the whole PN phase
of 20000 yrs, and between 10 and 50 considering the phase during which
the PN is bright enough to be observed at the detection limit of our
survey, typically several thousand yrs.  Taking into account that the
central region $\sim$4\arcmin $\times$4\arcmin\ is covered by a large
complex of \hii\ regions and the large foreground extinction in the
direction of \ic10, the number of 16 candidate PNe is  consistent 
with the expected population size.  

As shown in Figure~2, the number of our candidate PNe in \ic10\ is
also only slightly smaller than expected by scaling its luminosity
with that of the LMC, and normalizing with the known number of LMC PNe
where the survey for PNe is the most complete. The expected detectable
PN population for a galaxy with the V-band luminosity of \ic10\ would
in fact be  $\sim$30 PNe.  Our survey might therefore be slightly
less complete owing to the greater distance of \ic10 compared to the
LMC, the large reddening, and the presence of the large central
complex of
\hii\ regions.  Particularly, the star formation regions contribute 
considerably to the V-band luminosity, but they are too young to 
produce planetary nebulae.  Our method for estimating the number of 
PNe in a galaxy makes use of the V-band luminosity, without 
considering that the galaxy luminosity must depend on the star 
formation history.  The `effective' V-band luminosity is thus lower 
than the total V-band luminosity and consequently the dashed line in 
Figure~2 slightly overestimates the expected number of PNe in \ic10. 
   
We have also analyzed the 'incompleteness' of our survey as a function 
of m$_{\rm [O~III]}$ and of the distance to the centre of the galaxy, 
as described in M02.  We estimated the number of `missing' PNe by 
adding 'artificial stars' with various m$_{\rm [O~III]}$, as expected 
for the PN population of \ic10, in both \oiii\ and \stry\ frames.  The 
incompleteness (i.e a recovery rate of artificial stars less than 
50\%, Minniti \& Zijlstra \cite{minniti}) is due to the probability of 
missing an object in the \oiii\ or \ha+\nii\ images and/or to the 
probability of a wrong identification of a star in the continuum 
images.  We found that our survey of \ic10\ is incomplete for 
emission-line objects with m$_{\rm [O~III]} > 24.5$ and there is a 
probability of $\sim$5\% to find a star on the same line of sight and 
consequently to miss the emission-line object.  The loss of faint 
emission-line objects is quite uniform across the whole field, whereas 
the probability of overlapping with a star or with a large \hii\ 
region increases towards the centre, particularly in 
\ha+\nii\ images where it reaches 40-50\% within 4\arcmin\ from the    
centre.  Considering the fraction of the total mass inside the central   
4\arcmin$\times$4\arcmin\ region, we conclude that at least 5 PNe   
brighter than the completeness limit may have been missed there.   
    
    
Previous estimates of the distance to \ic10\ locate it to a distance   
ranging from 0.5 to 3 Mpc (Roberts \cite{roberts}; de Vaucouleurs \&   
Ables \cite{devau}; Wilson et al. \cite{wilson}; Tikhonov   
\cite{tikhonov}; Sakai et al. \cite{sakai99}).   
The planetary nebulae luminosity function (PNLF) is widely used as an
extragalactic distance indicator (Jacoby \cite{jacoby89}). When the PN
population size is small, the absolute magnitude of the bright cut-off
(-4.53 for a large sample of PNe, Ciardullo et al. \cite{ciardullo})
of the luminosity function increases because the brightest PNe could
not be observed (M\'endez et al. \cite{mendez}). In this case the PNLF
cannot be used as a distance indicator.  The m$_{\rm [O~III]}$ of the
brightest PN can give, however, an upper limit to the distance of
\ic10, $\sim$1.8~Mpc.   
In Sect. 5 we will discuss how the metallicity correction applies to 
the m$_{\rm [O~III]}$ of the brightest PN. 
   

  
\subsection{Planetary nebulae in \leoa}    
    
Our observations identified one candidate PN in \leoa\ (see Table 1
and Figure~1).  The survey by JL81 allowed the discovery of two
candidate PNe, which are shown in Figure~2 of their paper. These two
candidate appear clearly both in our continuum and emission-line
frames, therefore we conclude that they are normal stars.  During an
\ha\ survey of \leoa, Strobel et al. (\cite{stroebel})
have detected three \hii\ regions and an unresolved emission region that they
identified as a PN. This object corresponds to our candidate PN.  A
spectrum of this object was obtained by Skillman et
al. (\cite{skillman}), confirming its nature as a PN. From our \oiii\
and \ha+\nii\ fluxes, corrected for the foreground extinction
E(B-V)=0.02 (Strobel et al. \cite{stroebel}), we computed an
excitation index $R$=$1$, typical of a relatively low excitation PN, and/or
of low oxygen abundance. 
This is lower than the value $R$=$1.4$ computed using the line fluxes
in the spectrum by Skillman et al. (\cite{skillman}). Note however that
their data would also measure an unrealistic \ha/\hb\ ratio of 2.3,
suggesting that line fluxes in the blue region are overestimated. In
fact, adjusting their line fluxes in the blue to achieve the
theoretical value \ha/\hb=2.85, \oiii\ would be scaled so as to
provide $R$=$1.1$$\pm$$0.2$, in good agreement with our value.
   
Using eq.~1, the expected PN population of \leoa\ is 1 to 3 PNe, thus
consistent with the single PN observed.  The upper limit to the
distance modulus to \leoa\ given by the m$_{\rm [O~III]}$ of the PN,
without correcting for its low metallicity, is $\sim$26.3 (1.8~Mpc), very
close to the distance obtained by Sandage (\cite{sandage}) using the
three brightest stars in \leoa, but both methods are not reliable for
galaxies with such a small population size.  Our survey is fairly
complete up to m$_{\rm [O~III]}$=24.5 and no PN brighter than 24.5 mag
remain to be discovered.  Further deep spectroscopic study, as the one
by Skillman et al. (\cite{skillman}) who measured the O/H abundance,
will give fundamental information about the metal content in one of
the lower metal abundance galaxies of the Local Group.
    
\subsection{Planetary nebulae in  \sexa}    
    
Several emission-line objects are shown in the \oiii\ and \ha+\nii\
continuum subtracted images of \sexa.  The extended emission-line
sources are \hii\ regions previously studied by Hodge et al.
(\cite{hodge}), Dohm-Palmer et al. (\cite{dohm}) and supergiant
ionized filaments studied by Hunter \& Gallagher (\cite{hunter}).  We
identified one unresolved emission-line source present both in the
\oiii\ and \ha+\nii\ continuum-subtracted images (see Table~1 and Figure~1).  
This   candidate PN was previously discovered by JL81, and the
\oiii\ flux  presented in their paper is in agreement,    
within the errors, with our measure.  The R ratio for this candidate
PN, computed after correcting for reddening, E(B-V)=0.015 (vdB00), is R=0.5, indicating that
either that it is a low-excitation PN, a compact \hii\ region or that it has low
oxygen abundance, as we can expect from the low metallicity of \sexa\
(12+log(O/H)=7.49 from Skillman et al. \cite{skillman}).  It lies near
to the centre of the galaxy and has m$_{\rm [O~III]}$=22.46 which sets
an upper limit to its distance modulus of 26.7 (2.5~Mpc), without
correction for low-metallicity.  The completeness of our survey is up
to m$_{\rm [O~III]}$=24.5. No PN brighter than the 24.5~mag remain to
be discovered.
    
\sexa\ and \sexb\ have similar distances and are separated    
on the sky by only 10\fdg4.  Considering also their small velocity
difference (23 $\pm$6 km s$^{-1}$, vdB00), it is quite possible that
they had formed together and then drifted apart over a Hubble time
(vdB00).  Their V-band luminosities and their mass are also similar,
and thus their expected PN populations would be alike.  Five PNe were
discovered in \sexb\ (M02), while only one candidate PN is detected in
\sexa.  Statistically, this difference is only marginally significant,   
but may suggest some differences in their star formation history. In 
fact, M02 argued that \sexb\ has a large number of intermediate-age 
stars formed in the last 5 Gyrs, while \sexa\ has a stronger main 
sequence population (low- and intermediate-mass stars, Dohm-Palmer 
et al. \cite{dohm}) likely due to a stronger recent star formation. 
 
\section{Planetary nebulae in the Local Group} 
 
A total number of $\sim$2500 extragalactic PNe have been discovered
during the last thirty years in almost all the LG galaxies whose total
luminosity implies a population size large enough to allow the
presence of PNe ({\sl cf.}  Ford et al. \cite{ford}, Magrini et
al. \cite{magrini02}).
 
\subsection{The PN population size} 
 
Figure~2 shows the number of PNe discovered to date in galaxies of the
LG or in nearby groups, versus the V-band luminosity in solar units.
This is an updated version of Fig.~2 in M02 and includes the PNe
discovered in this paper (16 in \ic10\ and 1 in \sexa), as well as 19
PNe in NGC~6822 confirmed spectroscopically by P. Leisy (in preparation), 
1284 candidate PNe in M~31 (D. Carter, private
communication, see also Ford et al. \cite{ford}), one PN in WLM
(Zijlstra, private communication),  2389 PNe in the Milky Way
(Acker et al. \cite{acker}), 2 in IC~1613 (Magrini et al., in preparation).  
Compared to the previous version of this
plot, the overall dispersion of data is reduced and a better
relationship between the number of PNe and the V-luminosity of the
galaxies can be seen.  M~31 and M~33 are rather short of known
objects, probably because of extinction effects and incompleteness in
the existing surveys.
 
\begin{figure}    
\centering   
 \includegraphics[width=9cm]{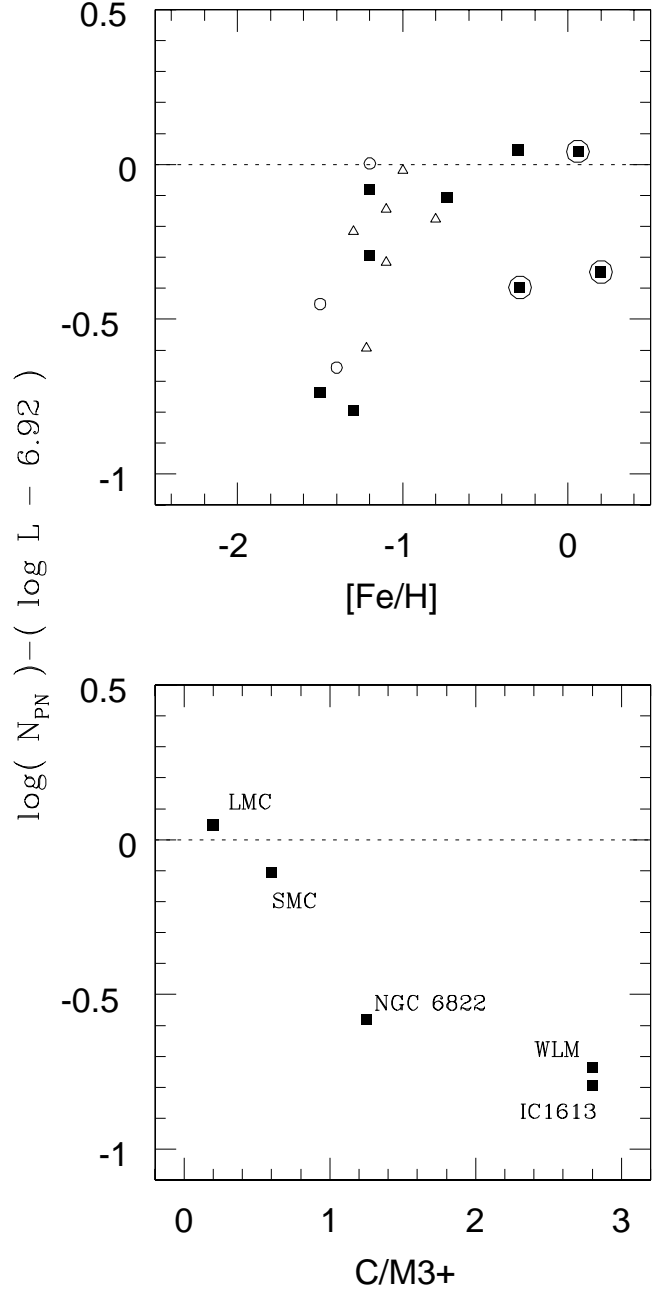}    
\caption{ 
The ratio between the observed and expected number of PNe ($\log
N_{\mbox{PN}}$ - ($\log(L_V)-6.92$), see text) as a function of the
metallicity of the galaxy (upper panel) or the ratio between carbon
and M stars (lower panel).   Squares are gas-rich local group galaxies,
encircled squares are the three main LG spirals (M33, the Milky Way
and M~31).  Triangles are gas-poor spheroidal types and circles the
galaxies probably not bound to the LG. In the lower panel only dwarf galaxies 
are shown.} 
\label{lgpn2}  
\end{figure}  
    
We have also investigated possible metallicity effects on the number
of PNe in a galaxy.  The upper panel of  Figure~3 shows the [Fe/H]
abundance of the galaxies vs. ($\log N_{\mbox{PN}}$ -
($\log(L_V)-6.92$)), which is the logarithm of the ratio between the
observed number of PNe ($N_{PN}$) and the expected population size
($\log (L_V)-6.92$, scaled from the data on \sexb\ where M02 found
that one PN is expected per L$_V$=10$^{6.92}$ solar luminosities).
The adopted number of PNe per unit luminosity is consistent with the
statement by vdB00 that 6 carbon (C) stars are found at M$_V$=-10,
making the expected ratio between PNe and C stars approximately 0.1.
Given that PNe are bright until they become optically thin at the
radius approximately of 0.05pc, which occurs after 2500~yr for a
nebular expansion velocity of 20~\kms, Figure~3 implies that
the carbon star life time must be ten times longer or
$\sim$25$\times$10$^3$~yrs, which is qualitatively correct.  Note that
in this paper we use two different relations for the number of PNe:
equation 1 and $\log(L_V)-6.92$.  Equation~1 derives the total number
of expected PNe, while $\log(L_V)-6.92$ is based on the observations
(so excluding faint objects, confusion, etc.) and thus infers the
expected observable number of PNe. Note that adopting a value
different from $6.92$ as the normalization factor would not affect the
discussion, as it would only shifts the relation $\log(L_V)-6.92$ up
or down.

Figure~3 shows that there is a slight tendency for the number of
detected PN to decrease with metallicity, if M~31 and M~33, whose
total observable population size in uncertain, are excluded (encircled
squares at bottom).  In this graph, we have also excluded galaxies
where the expected number of PNe is $<$1 PN (\leoa\ and Pegasus) and
IC~1613 which still lacks a proper survey for PNe.   
In particular, in Figure~3 there are some hints of a shortfall in the
observed number of PNe for [Fe/H]$<<$-1.0.  As [Fe/H]=-1.0 corresponds
to the point where the AGB wind is expected to be driven no longer by
dust, but only by pulsations (Zijlstra \cite{zijlstra99}), the lack of
PNe might suggest that below this point the PN formation rate is
largely reduced.  Note that Mira variables are only seen in globular
clusters when [Fe/H]$>$-1, suggesting there is significant change in
AGB evolution at this point.

The bottom panel of Figure~3 shows the PN deficit against the ratio
between carbon and M stars ($C/M3+$) for the galaxies where this
information is available, taken from Cook et al. (\cite{cook}). The
number of M stars here is those with spectral type M3 or later. This
number increase rapidly with lower metallicity. The trend in the
figure confirms the suggestion in the top panel.

As a matter of caution, it should however be noticed that part of the
deficit of detected PNe in metal poor galaxies might also depend on
the technique that is generally used to find PNe, i.e. \oiii\ 500.7~nm
imaging, as the \oiii\ emission decreases for low abundances of this
element, as shown in the LMC and SMC by Leisy (2003, in preparation).
On the other hand, our detection technique make use in addition of 
\ha\ + \nii\ images. 
We have considered also low-$R$ candidate, 
as the one discovered in \sexa, thus reducing the probability to lose PNe in 
low-metallicity galaxies.

 
\subsection{On the brightest PNe of the Local Group} 
 
The first studies of the PNLF showed no strong evidence for
metallicity dependence (Jacoby et al. \cite{j88}, Jacoby et
al. \cite{j90}).  Theoretical models by Jacoby (\cite{jacoby89})
suggested that the dependence of the bright cutoff of the PNLF with
metallicity is modest, $\propto Z^{0.5}$.  This effect is relatively
unimportant when metallicity differs from galaxy to galaxy by 30\% or
less, but it could lead to significant differences when a wider range
of metallicities is considered, as in the case of the LG.  Dopita et
al. (\cite{dopita}) examined the effects of metallicity on the
luminosity of PNe, modeling the variation of the \oiii\ magnitude
M$_{\rm [O~{III}]}$ with the oxygen abundance as
\begin{equation}   
\Delta M_{\rm [O~{III}]} =0.928{\rm[O/H]^2}+0.225{\rm[O/H]}+0.014, 
\end{equation}    
assuming the solar abundance of the oxygen to be 12+log(O/H)=8.87
(Grevesse et al. \cite{grevesse}). Eq. 2 indicates that M$_{\rm
[O~{III}]}$ is brightest when the PN oxygen abundance is near to
solar, while it decreases for both metal-poor and metal-rich galaxies.
This has been observed in relatively metal-poor galaxies (SMC,
12+log(O/H)=8.03; NGC~300, 8.35) by Ciardullo et al.
(\cite{ciardullo}), but it has never been observed in high metallicity
galaxies because in those galaxies solar-metallicity stars are also
available and they define the bright cutoff of the PNLF.
 
\begin{figure}   
\resizebox{\hsize}{!}{\includegraphics{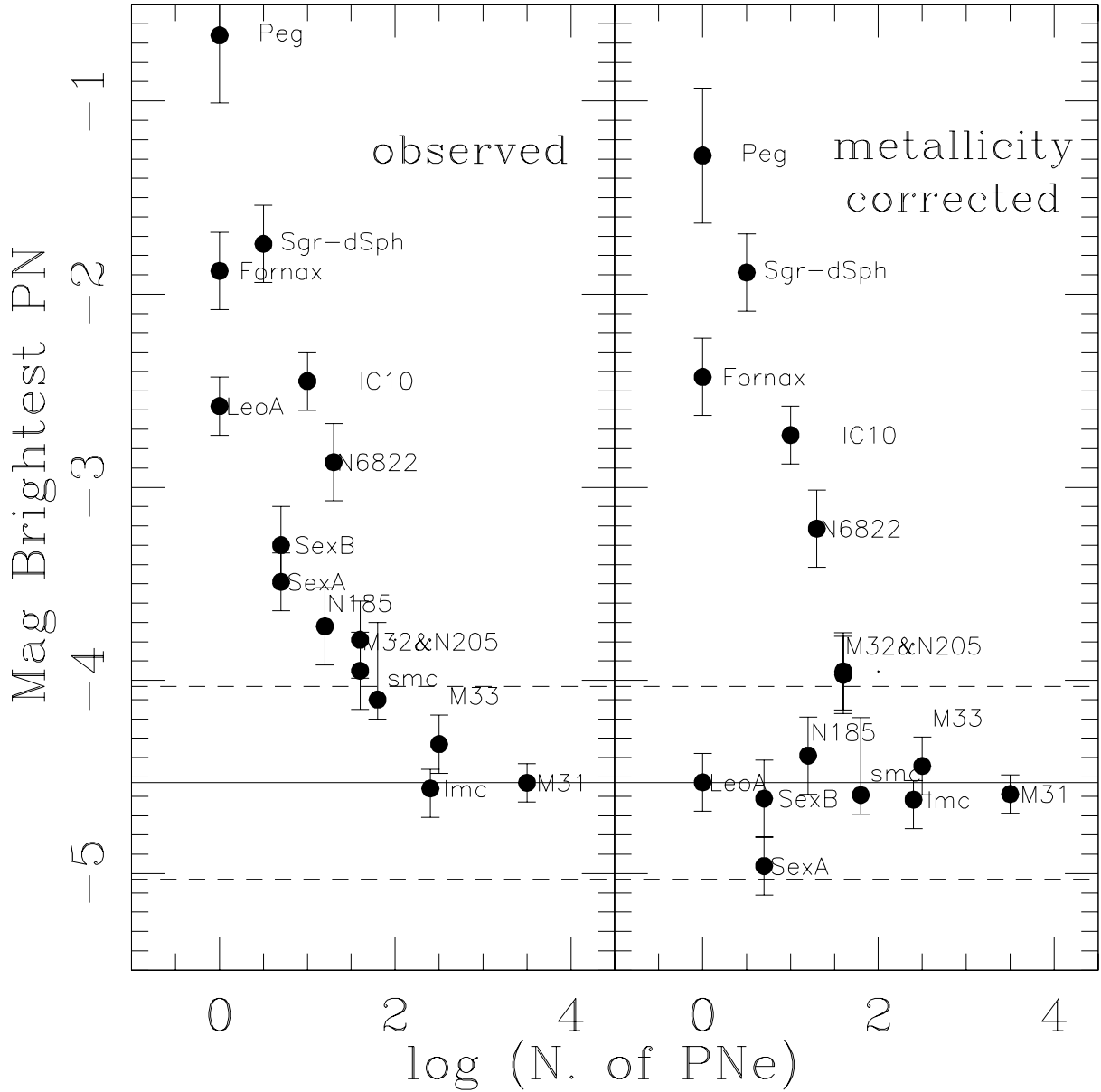}} 
\caption{The expected number of PNe vs the absolute  
magnitude (M$_*$) of the brightest PN, before (panel on the left) and
after (panel on the right) metallicity correction following Dopita et
al. (\cite{dopita}) prescription. For M~31, LMC and SMC we report the
magnitude of the cutoff of their PNLFs with errors from Ciardullo et
al. (\cite{ciardullo}). For M~33 the magnitude of the cutoff is from
Durrell et al. (\cite{durrell}). The horizontal line show the
magnitude of the PNLF cutoff M$_*$=-4.53 as expected for a galaxy with
a large population size.  Dashed lines mark 1-mag around M$_*$=-4.53.}
\end{figure}  
 
The correction given in equation 2 is significant when the metallicity
is much lower than solar. The LG is a good environment in which
investigate this issue, as oxygen abundance spans over a large range:
from 7.30 of \leoa\ to 9.0 of M~31 (from vdB00).  With the data
presented in this paper and in the previous surveys (Danziger et
al. \cite{danziger}, JL81, Killen \& Dufour \cite{killen}, Ciardullo
et al. \cite{ciardullo89}, Morgan \cite{morgan}, Zijlstra \& Walsh
\cite{zijlstra}, Leisy et al. \cite{leisy}, M02, Durrell et al. \cite{durrell}), an extensive
sample of the brightest planetary nebulae of the LG is available.  In
Figure~4, we present the PN population size versus M$_*$ or the cutoff
of the PNLF (when available, i.e. M~31, M~33, SMC, LMC, {\sl cf.}
Ciardullo et al. \cite{ciardullo}), Durrell et al. \cite{durrell}
before and after correcting for metallicity using eq.~2. Distances
used to compute the absolute magnitudes are from vdB00, observed
magnitudes are from the above cited papers.  The solid horizontal line show the magnitude of
the PNLF cutoff as expected for a galaxy with a large population size
and Solar metallicity (M$_*$=$-4.53$, Ciardullo \cite{ciardullo}).
 
After correcting for metallicity using eq.~2, a number of galaxies of 
the LG, especially those with a PNe population size larger than 10$^2$, 
are tightly grouped around the ``universal'' value M$_*$=$-4.53$.  This 
seems to confirm that the metallicity correction by Dopita et 
al. (\cite{dopita}) is at least qualitatively correct, and that 
the effects of metallicity are non-negligible for galaxies in the LG. 

For galaxies with a population size of 10$^2$ or less, 
M$_*$ is spread over a large range; it is likely that in these 
galaxies we start seeing the effect of having a small population size, 
as discussed in Sect.~4.1.2, together with some other effects, like 
for instance incompleteness of the existing surveys due to extinction 
or to the existence of areas of high stellar density or with large 
systems of \hii\ regions ({\sl cf.} e.g. with Sect.~4.1.1. for \ic10) .  In 
this respect, it is somewhat surprising to find small galaxies 
like \leoa, \sexb, and \sexa\ near M$_*$=$-4.53$, in spite of their 
very small PN population size.  
 
The corrected magnitudes of the brightest PNe allow us to revise our
upper limits to the distances of these galaxies presented previously
in this paper and in M02.  After
correcting for metallicity, the upper limits to the distance moduli to
\sexb\ would be 25.5 (1.3~Mpc), 25.4 (1.2~Mpc) for \sexa, and 24.3
(0.7~Mpc) for \leoa, in good agreement with other estimates (1.32 and
1.45 from vdB00, 0.8 from Dolphin et al. \cite{dolphin}),
respectively, whereas for \ic10\ the correction is not remarkable. We
stress once more, however, that caution should be used when
determining distance for such low-luminosity systems using PNe.

\section{Summary and conclusions}    
    
In this paper, we have presented the discovery of sixteen new
candidate PNe in \ic10, and confirmed two previously known candidate
PNe, one in \leoa\ and the other one in \sexa.  These observations,
together with other ones that will be presented in the future for
other LG galaxies, provide a further improving in our understanding of
the PN population of the Local Group.  The behaviour of the numbers of
PNe with galaxy metallicity has been discussed, finding a possible
lack of PN when [Fe/H]$<<$-1.0.  The magnitudes of the brightest PNe
of the LG have been reviewed and the possibility to measure distance
using the magnitude of the brightest PN, after correcting for
metallicity, has been discussed.
   
Future spectroscopic studies of individual objects will confirm
their nature as PNe and allow detailed chemical abundance studies. 
 In addition, they will provide unique information
about the star formation history and chemical evolution of the parent
galaxies, using tracers of stellar populations of intermediate to old
age, and thus fully complementary to the information obtained from
older (e.g. population II red giants) and younger (e.g. HII regions)
classes of objects.
    
\begin{acknowledgements}   
The data have made publically available through the Isaac Newton Groups' 
Wide Field Camera Survey Programme. 
This research has made use of the NASA/IPAC Extragalactic Database (NED),  
the APM and USNO-A2.0 Sky Catalogues, and the  ESO Online Digitized Sky Survey. 
We would like to thank an anonymous referee for  comments and suggestions which 
improved our paper. 
\end{acknowledgements}

\end{document}